\documentclass[fleqn,usenatbib]{mnras}

\usepackage{newtxtext,newtxmath}

\usepackage{graphicx}
\graphicspath{{figure/}}

\defcitealias{Madau2014}{MD14}
\defcitealias{Bruzual2003}{BC03}
\defcitealias{Yung2024}{Y24a} 
\defcitealias{Yung2024a}{Y24b} 

\usepackage[T1]{fontenc}

\DeclareRobustCommand{\VAN}[3]{#2}
\let\VANthebibliography\thebibliography
\def\thebibliography{\DeclareRobustCommand{\VAN}[3]{##3}\VANthebibliography}

\usepackage{graphicx}	
\usepackage{amsmath}	
\usepackage[labelformat=simple]{subcaption}
\graphicspath{{figures/}}
\usepackage{url}
\usepackage{multirow}

\newcommand{\msun}{\;M$_\odot$}

\newcommand{\gureft}{\textsc{gureft}}

\usepackage{xcolor}

\title[$\Lambda$CDM is still not broken]{$\Lambda$CDM is still not broken:  empirical constraints on the star formation efficiency at \boldmath{$z\sim$}12 -- 30}

\author[L. Y. A. Yung, R. S. Somerville, K. G. Iyer]{L. Y. Aaron Yung,$^{1}$\thanks{e-mail: yung@stsci.edu}
Rachel S. Somerville,$^{2}$\thanks{e-mail: rsomerville@flatironinstitute.org}
and Kartheik G. Iyer$^{3}$\thanks{Hubble Fellow}\\
$^{1}$Space Telescope Science Institute, 3700 San Martin Drive, Baltimore, MD 21218, USA\\
$^{2}$Center for Computational Astrophysics, Flatiron Institute, 162 5th Ave, New York, NY 10010, USA\\
$^{3}$Columbia Astrophysics Laboratory, Columbia University, 550 West 120th Street, New York, NY 10027, USA\\
}

\date{Accepted XXX. Received YYY; in original form ZZZ}

\pubyear{\the\year{}}

\begin{document}
\label{firstpage}
\pagerange{\pageref{firstpage}--\pageref{lastpage}}
\maketitle

\begin{abstract}
The James Webb Space Telescope continues to push back the redshift frontier to ever earlier cosmic epochs, with recent announcements of galaxy candidates at redshifts of $15 \lesssim z \lesssim 30$. We leverage the recent \gureft\ suite of dissipationless $N$-body simulations, which were designed for interpreting observations in the high redshift Universe, and provide predictions of dark matter halo mass functions and halo growth rates for a state-of-the-art cosmology over a wide range of halo masses from $6 < z< 30$. We combine these results with an empirical framework that maps halo growth rate to galaxy star formation rate and then to rest-frame UV luminosity. We find that even if all of the photometrically selected $15 \lesssim z \lesssim 30$ galaxy candidates are real and actually at these extreme redshifts, there is no fundamental tension with $\Lambda$CDM, nor are exotic explanations required. With stellar light-to-mass ratios similar to those in well-studied lower redshift galaxies, our simple model can account for the observed extreme ultra-high redshift populations with star formation efficiencies that peak at values of 20-65 percent. Bursty star formation, or higher light-to-mass ratios such as are expected for lower metallicity stellar populations or a top-heavy Initial Mass Function, would result in even lower required star formation efficiencies, comparable to values predicted by high resolution numerical simulations of high-surface density star forming clouds.
\end{abstract}



\section{Introduction}

One of the primary science goals of the James Webb Space Telescope (JWST) is to probe the earliest galaxies in the Universe, which ended the Cosmic Dark Ages and brought about `Cosmic Dawn'. With its unprecedented infrared sensitivity, from the very first data release, JWST broke the existing `redshift barrier', and discovered multiple $z\gtrsim 10$ galaxy candidates \citep{Finkelstein2022a,Castellano2022,Bouwens2023b}. Over the past few years, the samples of $z\gtrsim 10$ galaxies have rapidly expanded with statistically robust studies of multiple, wider area fields \citep{Adams2022,Adams2024,Castellano2023,Donnan2022,Donnan2024,Finkelstein2023,Finkelstein2024,Leung2023a,Robertson2024,Whitler2025}. Many of these candidates have now been spectroscopically confirmed to lie at `ultra-high' redshift \citep{Harikane2023,ArrabalHaro2023,Curtis-Lake2023,Wang2023,Harikane2025,Kokorev2025}\footnote{In this paper we use the term `ultra-high redshift' or ultra-high-z to refer to galaxies at $z \gtrsim 10$.}. The number density of these ultra-high redshift galaxies is significantly in excess of both the expectations from extrapolations of lower redshift observations and empirical models, and the predictions of pre-launch physics-based models and simulations \citep[e.g.][]{Finkelstein2023,Leung2023a,Finkelstein2024,Adams2024}. 

The redshift frontier seems to be ever receding, with fairly robust samples of photometrically selected galaxy candidates (and a few spectroscopically confirmed objects) now identified out to $z\sim 14$ \citep{Perez-Gonzalez2023,Carniani2024a,Carniani2024,Robertson2024,Whitler2025}. Very recently, several works have announced the detection of photometrically selected galaxy candidates at even higher redshifts $z \sim 16$--25  \citep{Perez-Gonzalez2025,Castellano2025}. 

In the standard $\Lambda$CDM paradigm for cosmological structure formation, galaxies form within dark matter dominated halos \citep{Blumenthal1984}. Within this framework, the cosmological parameters are known to fairly high precision, and the initial conditions of the Universe are well constrained by observations of the Cosmic Microwave Background. Under the assumptions of `vanilla' $\Lambda$CDM, it is also straightforward to predict the number density of gravitationally bound dark matter halos as a function of their mass and as a function of cosmic time, and the growth rate of these halos, using dark matter only $N$-body simulations. One of the inescapable predictions of $\Lambda$CDM is that structure grows hierarchically over time via gravitational instability. Thus, the abundance of more massive halos increases over cosmic time \citep[e.g.][]{Mo2002}.

This is in apparent tension with the number density of UV luminous galaxies at $z\gtrsim 10$ discovered by JWST, which declines much more slowly than the predicted abundance of massive halos \citep[e.g.][]{Robertson2024}. Since the first science results on the early Universe from JWST were released, there have been claims that these observations might be in fundamental tension with standard $\Lambda$CDM \citep{Boylan-Kolchin2023,Lovell2022}, implying that the foundations of most existing work on galaxy formation and cosmology may need to be reconsidered. However, the difficulty is always that we do not know the masses of the halos that host the galaxies observed by JWST, as we do not know the efficiency with which these early galaxies were able to convert inflowing baryons into stars, or the amount of light per unit mass emitted by these stars in the UV. Proposed solutions that are much less dramatic than revising the fundamentals of $\Lambda$CDM include invoking a higher star formation efficiency at early times, perhaps due to weaker stellar or supernova feedback \citep{Dekel2023,Li2023,Somerville2025}, bursty star formation \citep{Shen2023a, Sun2023a, Gelli2024}, reduced dust attenuation \citep{Ferrara2023}, higher light-to-mass ratios in the UV due to lower metallicity stellar populations or a top-heavy IMF \citep{Trinca2024, Yung2024}, or a contribution to the UV from accreting black holes \citep{Trinca2024}. These revised models are apparently able to successfully account for the observed number density of UV-bright galaxies at $z \lesssim 12$, but it is less clear that they will be consistent with the more recent detections at $z\gtrsim 14$ \citep[e.g.][]{Whitler2025}. The announcement of the recently detected extreme ultra-high-z population at $z \sim 15$--30 prompts the question: \emph{if these objects are really at their proposed redshifts, would they be in fundamental tension with $\Lambda$CDM}, or require an exotic explanation such as accreting primordial black holes \citep{Matteri2025} or cosmic strings \citep{Koehler2024}?

In this paper, we adopt a very simple empirical model (various versions of which have been used in many previous works) that links dark matter halos and their growth rates to star formation and UV light. As a backbone, we leverage a recent suite of dark matter only $N$-body simulations which probe the ultra-high to extreme ultra-high-z Universe with unprecedented dynamic range and precision \citep{Yung2024a}. We first ask the question: if \emph{all} of the recently announced $z \sim 15$--30 galaxy candidates are real, and they are truly at their estimated redshifts, would this be in fundamental tension with $\Lambda$CDM? To answer this question, we assume that all inflowing baryons are instantly converted into stars, and that stellar populations have UV light-to-mass ratios similar to those in well-studied lower redshift galaxies ($z\sim 6$--10). Finding that this extreme model \emph{overproduces} the observed galaxy populations even out to $z\sim 25$, we then constrain the required efficiency of converting baryons to stars to explain the observations at face value, again using conservative assumptions about UV light-to-mass conversions. 

The structure of the paper is as follows. In Section~\ref{sec:methods:obs}, we briefly describe the observational constraints used in our study. In Section~\ref{sec:gureft}, we describe the $N$-body simulation suite that forms the backbone of our calculations. In Section~\ref{sec:empirical}, we describe the empirical modeling framework and our method for constraining its parameters using observational constraints. In Section~\ref{sec:results}, we show our predicted UV luminosity functions from $z\sim 12$--30 under the maximal star formation efficiency assumption, and then show the constraints on the star formation efficiency at these redshifts from the observed UV LFs. We discuss caveats of our analysis and implications of our results in Section~\ref{sec:discussion}, and conclude with a summary in Section~\ref{sec:conclusions}.

\section{Methods}
\label{sec:methods}

In this section, we provide a detailed description of the individual components used to construct the empirical model. Throughout this work, magnitudes are expressed in the AB system \citep{Oke1983} and all logarithms are base 10 unless otherwise specified.

\subsection{Summary of Observational Samples}
\label{sec:methods:obs}
We adopt observational rest-UV luminosity function constraints from a compilation of studies at redshifts $11.5 \lesssim z \lesssim 13$ ($\bar{z} \simeq 12$), $13 \lesssim z \lesssim 15$ ($\bar{z} \simeq 14)$, $16 \lesssim z \lesssim 20$ ($\bar{z} \simeq 17$) and $20 \lesssim z \lesssim 30$ ($\bar{z} \simeq 25$). We adopt only recent UVLF estimates from studies that incorporate the in-flight calibration of NIRCam (post February 2023). We include the luminosity function constraints from PRIMER \citep{Donnan2024}, EPOCHS (which combines data from PEARLS\footnote{\citealt{Windhorst2023}}, CEERS\footnote{\citealt{Bagley2023, Finkelstein2023, Finkelstein2024}}, GLASS\footnote{\citealt{Treu2022}}, NGDEEP\footnote{\citealt{Bagley2024}}, and the first data release of JADES\footnote{\citealt{Curtis-Lake2023, Robertson2023}};  \citealp{Adams2024}), JADES \citep{Robertson2024,Whitler2025}, CEERS \citep{Finkelstein2024}, the NIRCam parallel field from the MIRI deep survey in the Hubble Ultra Deep Field \citep{Perez-Gonzalez2023}, COSMOS-Web \citep{Casey2024, Franco2025}, the ASTRODEEP-JWST multi-band catalogs for the CEERS, Abell-2744, JADES, NGDEEP, and PRIMER fields \citep{Castellano2025}, MIDIS+NGDEEP \citep{Perez-Gonzalez2025}, and the PANORAMIC survey \citep{Weibel2025}.

\subsection{The GUREFT N-body simulation suite}
\label{sec:gureft}
In this work, we combine dark matter halo mass functions and halo growth rates from a suite of simulations that allows us to cover a broad range in halo mass from $6\lesssim z \lesssim 30$. The Bolshoi-Planck/MultiDark suite \citep{Klypin2016,Rodriguez-Puebla2016} were run with the Adaptive Refinement Tree (ART) code \citep{Kravtsov1997, Gottlober2009}. The \gureft\ suite \citep[][hereafter \citetalias{Yung2024a}]{Yung2024a} was run with \textsc{gadget}-2 \citep{Springel2005a}. All of these simulations adopt cosmological parameters that are consistent with one another and with the observational constraints from \citet{Planck2018}. The values of these parameters, which we adopt throughout this work, are as follows: $\Omega_\text{m} = 0.307$, $\Omega_\Lambda = 0.693$, $H_0 = 67.8$ km s$^{-1}$ Mpc$^{-1}$, $\sigma_8$ = 0.829, and $n_s = 0.96$.

In all simulations used here, halos are identified using the seven-dimensional phase-space halo finder \textsc{rockstar} \citep{Behroozi2013a} and halo accretion rates were computed using \textsc{consistent-trees} \citep{Behroozi2013b}. We adopt the \citet{Bryan1998} halo mass definition, which is defined relative to the cosmic critical matter density.

\subsubsection{Halo mass functions to ultra-high-z}
\begin{figure}
    \includegraphics[width=\columnwidth]{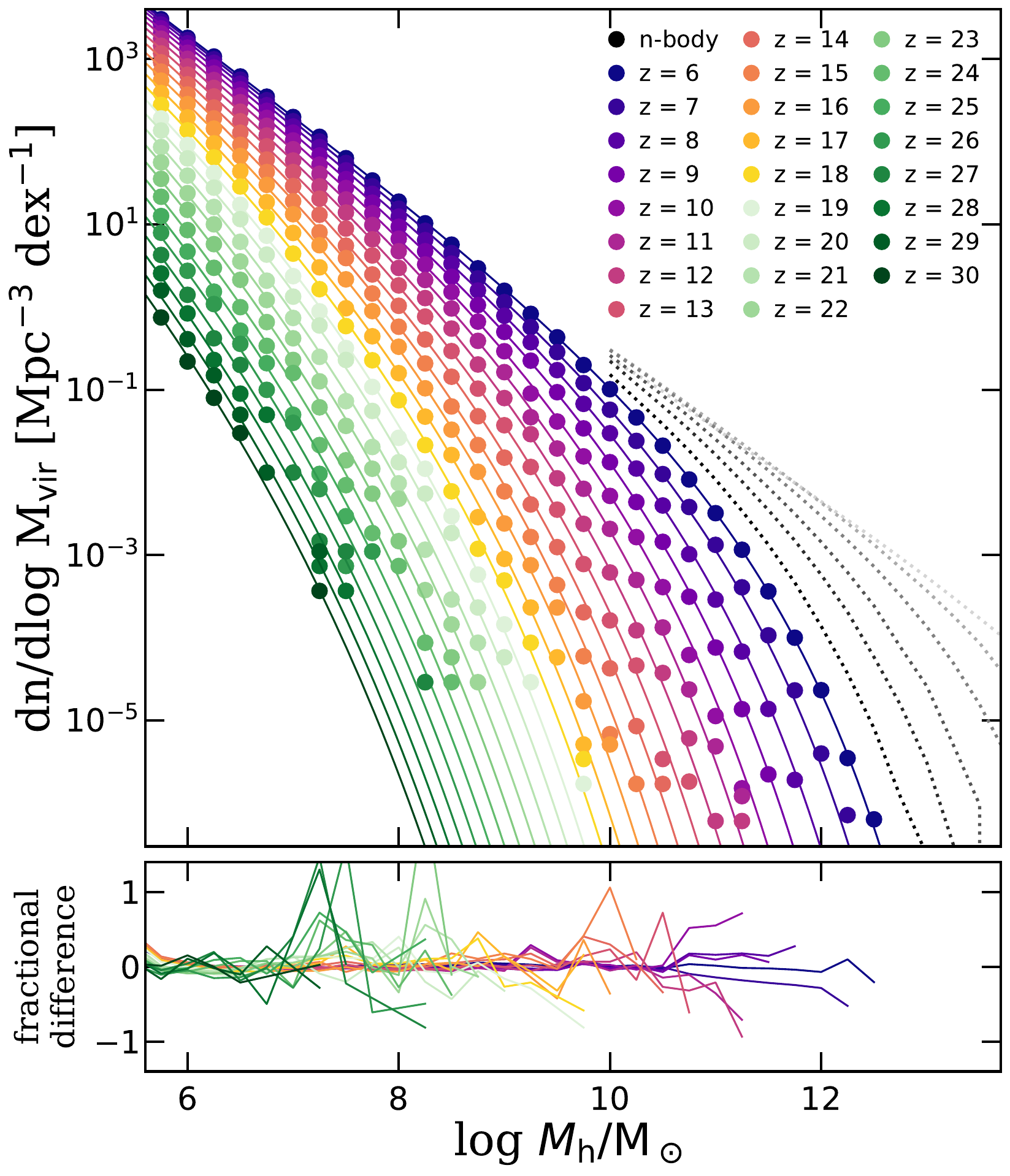}
    \caption{
        \textit{Top panel:} HMFs from $z=6$ to 30 (color coded as shown on the figure key) constructed with halos from the \gureft\ simulations and the VSMDPL simulation. Circular symbols show the binned data from the simulations and the results of the fitting function are shown with solid lines.  In addition, we show HMFs between $0 \lesssim z \lesssim 5$ from the Bolshoi-Planck simulation (gray dotted lines). Results are only shown for well-resolved halos that contain at least 100 particles. \textit{Bottom panel}: The fractional differences between the $N$-body HMFs and fitting functions $(\phi_\text{fitted} - \phi_{n-\text{body}})/\phi_{n-\text{body}}$.
    }
    \label{fig:fitted_hmf_z30}
\end{figure}

Joining together halos extracted from the suite of \gureft\ simulations and the Very Small MultiDark Planck (VSMDPL) simulation from the MultiDark suite, we construct halo mass functions (HMFs) back to $z\sim30$ and down to $M_\text{h} \sim 10^{5}$ \msun, as shown in Fig.~\ref{fig:fitted_hmf_z30}. We also show HMFs from the Bolshoi-Planck simulations at $0 \lesssim z \lesssim 5$. These are truncated below $M_\text{h} \sim 10^{10}$ \msun, which is representative of the typical mass resolution of large volume cosmological simulations.

We adopted the HMF fitting function from \citet{Rodriguez-Puebla2016}, which is modified from \citet{Tinker2008}:
\begin{equation}
    \frac{dn}{dM_\text{h}} = f(\sigma) \frac{\bar{\rho}_\text{m}}{M^2_\text{h}} \left| \frac{d \ln\sigma^{-1}}{d \ln M_\text{h}} \right| \text{,}
\end{equation}
where $\bar{\rho}_\text{m}$ is the mean matter density in the Universe. $\sigma$ is the amplitude of the perturbations, which we adopt the following fitting function
\begin{equation}
    f(\sigma) = A \left[ \left(\frac{\sigma}{b} \right)^{-a} + 1\right] e^{-c/\sigma^2} \text{,}
\end{equation}
where $\chi_i = A$, $a$, $b$, and $c$ are free parameters given by $\chi_i = \chi_{0,i} + \chi_{1,i} + \chi_{2,i}$, with the best-fitting parameters fitted to the \gureft\ + MultiDark HMFs between $z = 6$ to 30 presented in Table \ref{tab:HMF_fitting_param}.
Note that these have been revised from \citetalias{Yung2024a} after the inclusion of \gureft\ HMFs at $z > 19$.
Here, $\sigma\equiv \sigma(M_\text{h}) D(z)$, where $D(z)$ is the linear growth-rate factor and we adopt the following fitting function for $\sigma(M)$:
\begin{equation}
    \sigma(M_\text{h}) = \frac{26.8 y^{0.41}}{1+6.18 y^{0.23} + 4.64 y^{0.37}} \text{,}
\end{equation}
where $y\equiv1/M_\text{h,12}$ and $M_\text{h,12}\equiv M_\text{h}/(10^{12}\text{\msun})$.
We note that we are only providing fitting functions in physical units (without a little $h$ scaling) and this fitting function for $\sigma$ is adjusted to reflect that. 
While keeping the same functional form, we also fitted to a wider halo mass range than the one originally considered by \citet{Rodriguez-Puebla2016}, which diverges significantly from the exact $\sigma_m$ at $\log(M_\text{h}/\text{\msun}) < 10$, in order to accommodate the wider range of halo masses considered in this work. The updated fitting function presented here is capable of producing $\sigma({M_\text{h}})$ that has a fractional difference within $\pm0.01$ per cent across $6 \lesssim \log(M_\text{halo}/\text{\msun}) \lesssim 13$ when compared to $\sigma(M)$ computed with the cosmological calculator \textsc{colossus}\footnote{\url{https://bdiemer.bitbucket.io/colossus/}, version 1.3.5} \citep{Diemer2018}. See \citetalias{Yung2024a} for details. 
The fitted HMFs are also shown in Fig.~\ref{fig:fitted_hmf_z30}. Across these wide mass and redshift ranges, the fractional differences between our fitting functions and the $N$-body simulations are within an acceptable range (e.g. $< 5$ per cent), with a few exceptions occurring mostly at the massive end of the HMFs where there are relatively few halos in the simulated volumes. 

We note here that fitting functions that have been extrapolated from much lower redshift, and analytic models based on the Press-Schechter formalism \citep[e.g.][]{Sheth1999} can be quite inaccurate at these extreme redshifts. We show a comparison between our results and other commonly used HMF from the literature in Fig.~\ref{fig:hmf_compare_ultraz}. The discrepancies can be up to an order of magnitude, and vary with halo mass and redshift. 

We provide a jupyter notebook that computes the updated fitting functions presented here at \url{https://github.com/lyaaronyung/GUREFT}.

\begin{table}
	\centering
	\caption{Halo Mass Function fitting parameters}
	\label{tab:HMF_fitting_param}
	\begin{tabular}{cccc}
		\hline
        $\chi_i$ & $\chi_{0,i}$ & $\chi_{1,i}$ & $\chi_{2,i}$ \\
        \hline
		A & $0.21307778$ & $-0.01042236$ & $ 0.00013897$\\
        a & $0.94192066$ & $ 0.04453040$ & $-0.00202483$\\
        b & $3.27712602$ & $-0.01313422$ & $ 0.01027465$\\
        c & $1.15214631$ & $0.012866285$ & $-0.00065572$\\
		\hline
	\end{tabular}
\end{table}

\subsubsection{Halo mass accretion rates at ultra-high $z$}

The halo mass accretion rates adopted in this work are computed using the \textsc{consistent-trees} algorithm, which constructs halo merger trees by tracking the growth of halos and establishing progenitor-descendant links across cosmological simulation snapshots. Descendant halos are identified based on a particle-based algorithm, with additional steps to account for positions and velocities that change across snapshots under the influence of gravity, reject spurious descendants, and insert filler halos when needed. In order to mitigate numerical noise from snapshot-to-snapshot fluctuations, we use halo accretion rates that are averaged over a halo virial dynamical time, defined as $t_\text{dyn}(z) = [(4/3)\;\pi\;G\rho_\text{vir}(z)]^{-1/2}$, where $\rho_\text{vir}$ is the \citet{Bryan1998} virial mass definition.

\begin{figure}
    \includegraphics[width=\columnwidth]{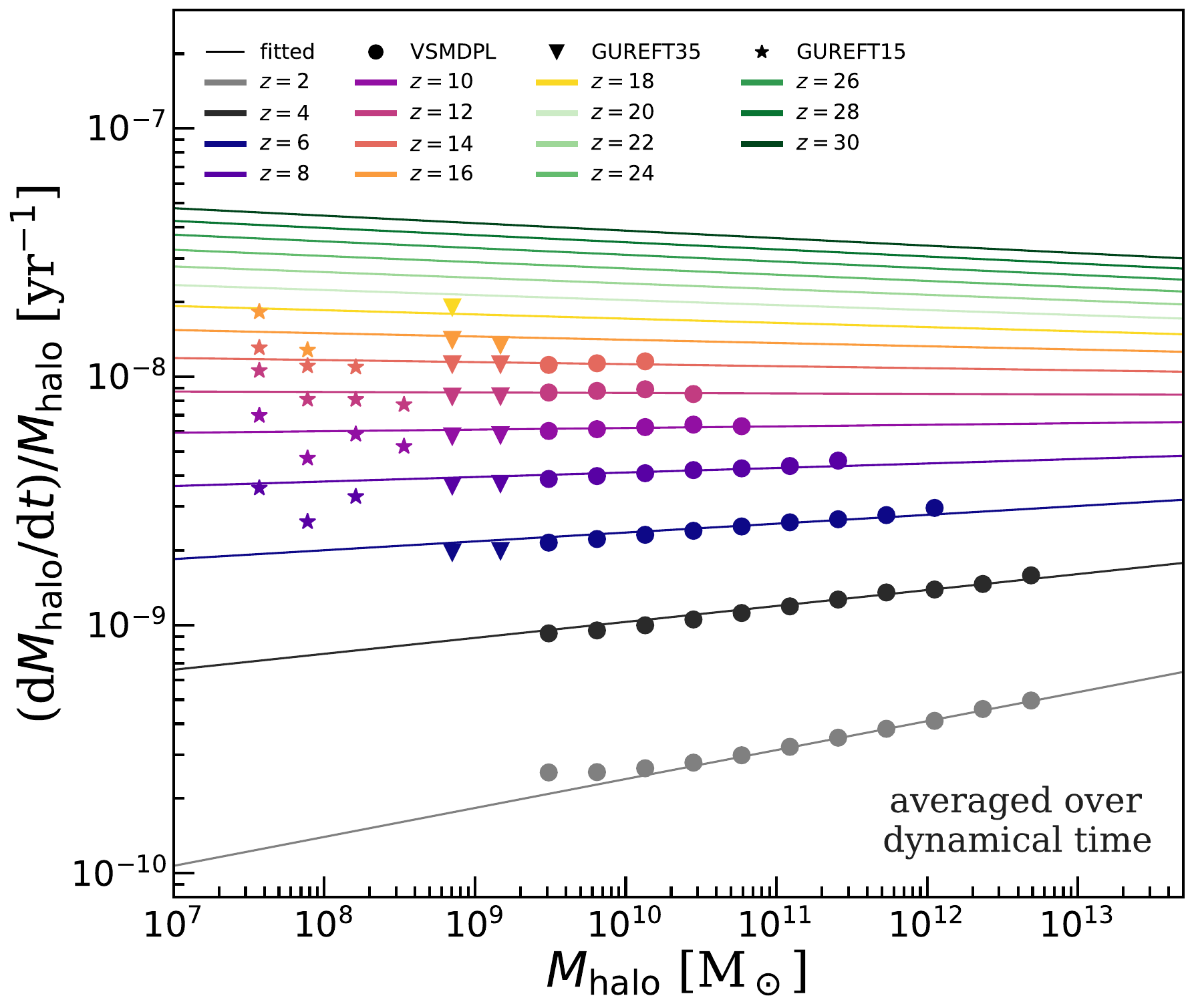}
    \caption{
        Halo mass accretion rate normalized to halo mass, $(dM_\text{h}/dt)/M_\text{h}$, averaged over a dynamical time, as a function of $M_\text{h}$ across a wide range of redshifts. The data points show results extracted from $N$-body simulations, including VSMDPL (circles), \gureft-35 (triangles), and \gureft-15 (stars). 
        The solid lines show our fitting functions, which are obtained by fitting to data from $z = 2$ to 18. The results of extrapolating the fitting functions to $z = 30$ are also shown. Both the data points and lines are colour-coded by redshift as shown in the figure key.
    }
    \label{fig:fitted_dMhdt_Mh}
\end{figure}

Fig.~\ref{fig:fitted_dMhdt_Mh} shows halo accretion rates averaged over one dynamical time from VSMDPL, \gureft-35, and \gureft-15. We fit these using the functional form:
\begin{equation}
    dM_\text{h}/dt(M_\text{h},z) = \beta(z)(M_\text{h,12}E(z))^{\alpha(z)}
\end{equation}
and find the parameters $\alpha(z)$ and $\beta(z)$ are well fit by:
\begin{equation}
\begin{split}
    \alpha(z) &= 0.948 + 0.694 a - 0.565 a^2\\
    \log_{10}\beta(z) &= 2.673 - 2.075 a + 0.891 a^2 \text{,}
\end{split}
\end{equation}
where $M_\text{h,12} \equiv M_\text{h}/(10^{12} \text{\msun})$ and $a = 1/1(1+z)$ is the scale factor. Note that these are revised from the results presented in \citetalias{Yung2024a} due to the inclusion of additional data $z>15$ and the inclusion of VSMDPL. Once again, we caution that fits extrapolated from much lower redshifts can be inaccurate (see \citetalias{Yung2024a}) and note that this fitting function is not intended for use at $z < 2$. We have verified that these results are insensitive to the exact choice of averaging timescale. 

\subsection{Empirical model framework}
\label{sec:empirical}

Our simple model is based on the ansatz that the star formation rate of a galaxy is proportional to the growth rate of its host halo, times the universal baryon fraction, times a halo mass and redshift dependent efficiency factor \citep{Tacchella2018,Shen2023a,Shuntov2025}: 
\begin{equation}
    \text{SFR} = \epsilon_*(M_h, z)\, f_\text{b} \dot{M}_\text{\rm h}
    \label{eqn:sfr}
\end{equation}
where $f_b=0.1573$ is the universal baryon fraction for our chosen cosmology, $\dot{M}_\text{\rm h}$ is the halo growth rate averaged over one halo dynamical time, and 
\begin{equation}
\epsilon_{\rm *} = \frac{2 \epsilon_0}{(M_h/M_0)^{-\alpha} + (M_h/M_0)^{\beta}}
\label{eqn:epsilon_star}
\end{equation}
where $\epsilon_0$, $M_0$, $\alpha$, and $\beta$ are free parameters.

We note here that throughout this work we use the terms star formation efficiency and baryon conversion efficiency interchangeably, and we define this quantity (denoted as $\epsilon_*$) as the ratio of the current SFR to the rate that baryons are flowing into the halo, as described above. Both of these quantities imply a timescale over which they are measured, and we implicitly assume that the relevant timescale for the SFR is that on which most of the UV light is produced, and the latter is the halo dynamical time. Based on the star formation histories of $z\gtrsim 10$ galaxies predicted by the physics-based Santa Cruz semi-analytic models run within \textsc{gureft} merger trees \citep{Yung2024,Somerville2025}, the halo dynamical timescale, which gradually rises from a few tens of Myrs at $z > 20$ to $\sim100$ Myr at $z \sim 10$, approximately matches the period over which young stellar populations dominate the UV emission (Yung et al. in prep). Our definition of $\epsilon_*$ is equivalent to the alternate, commonly used definition of baryon conversion efficiency $m_*/(f_b M_{\rm h})$ only if $\epsilon_*$ does not change with time, which is manifestly in conflict with the assumed halo mass and redshift dependence of $\epsilon_*$ in our model. It is also different from the star formation efficiency definitions often quoted in studies of star formation on molecular cloud scales, which are typically the fraction of gas in the initial cloud that is converted into stars over the lifetime of the cloud, or over one freefall time.

We then assume a fixed conversion factor from SFR to UV luminosity $L_{\rm UV}$:
\begin{equation}
    L_\text{UV} = \text{SFR} / \mathcal{K}_\text{UV}
\end{equation}

where we adopt $\mathcal{K}_\text{UV} = 0.72 \times10^{-28}$ M$_\odot$ yr$^{-1}$ erg$^{-1}$ s Hz from \citet[][hereafter MD14]{Madau2014}, converted to a Chabrier \citep{Chabrier2003a} stellar initial mass function. We discuss the basis for this choice and the implications of other choices in Section~\ref{sec:discussion}. With this choice, we implicitly assume that there is no dust attenuation in the rest-UV.
$L_\text{UV}$ is converted to $M_\text{UV}$ in the AB magnitude system using the standard conversion $\log_{10}(L_\text{UV}/(\text{erg}\;\text{s}^{-1} \text{Hz}^{-1})) = 0.4(51.63-M_\text{UV})$.

\subsection{Fitting Procedure}
\label{sec:fitting}
\begingroup
\renewcommand{\arraystretch}{1.5}
\begin{table}
\centering
\caption{Median values and 1$\sigma$ uncertainties on the parameters in equation \ref{eqn:sfr} for the four redshift bins. }
\begin{tabular}{ccccc}
\hline
redshift & $\epsilon_0$ & $M_0$ & $\alpha$ & $\beta$\\
\hline
$11.5 < z < 13$ & $0.46_{-0.29}^{+0.36}$ & $10.94_{-0.69}^{+0.58}$ & $1.17_{-0.46}^{+0.53}$ & $0.41_{-0.28}^{+0.32}$ \\
$13 < z < 15$ & $0.40_{-0.21}^{+0.38}$ & $10.40_{-1.05}^{+0.88}$ & $0.97_{-0.54}^{+0.67}$ &  $0.39_{-0.27}^{+0.32}$\\
$16 < z < 20$ & $0.69_{-0.30}^{+0.22}$ & $10.02_{-0.33}^{+0.75}$ & $1.28_{-0.65}^{+0.52}$ & $0.36_{-0.25}^{+0.33}$\\
$20 < z < 30$ & $0.64_{-0.26}^{+0.25}$ & $8.53_{-0.42}^{+1.26}$ & $0.97_{-0.69}^{+0.73}$ & $0.42_{-0.29}^{+0.31}$\\
\hline
\end{tabular}
\label{tab:epsilonstarparams}
\end{table}
\endgroup

We constrain the free parameters of our model ($\epsilon_0$, $M_0$, $\alpha$, $\beta$) using Markov Chain Monte Carlo (MCMC) with the \texttt{emcee} package \citep{Foreman-Mackey2013}. We adopt uniform priors  ($7<\log(M_0/$\msun$)<12$, $0.01<\epsilon_0 < 1.0$, $0<\alpha<2$, $0<\beta<0.9$) for the likelihood function, where $\epsilon_*=1.0, \alpha=\beta=0$ corresponds to the maximally efficient scenario. For the likelihood function, we adopt a hybrid approach that accounts for both observational uncertainties and upper limits from non-detections, which need to be accounted for to robustly model the posteriors (especially at the highest redshifts). The combined likelihood is a sum of the two contributions, 
$\mathcal{L} = \mathcal{L}_{\rm det} + \mathcal{L}_{\rm lim}$, following a treatment of upper limits as described in \citet{Sawicki2012} and \citet[][see appendix D for a derivation]{Ishikawa2022}. We run the MCMC with 32 walkers for 10,000 steps, discarding the first 100 steps as burn-in and thinning by a factor of 15 to reduce autocorrelation. The resulting posteriors are shown in Fig.~\ref{fig:efficiency_param_posteriors}, and the medians and 1$\sigma$ values of the parameter posteriors are provided in Table~\ref{tab:epsilonstarparams}. It is important to note that while the posteriors span a range of values and significant uncertainties prevent us from ruling out the maximal case, it is a clear indication that a range of physically plausible scenarios are consistent with the observed data.  

\begin{figure*}
    \includegraphics[width=2\columnwidth]{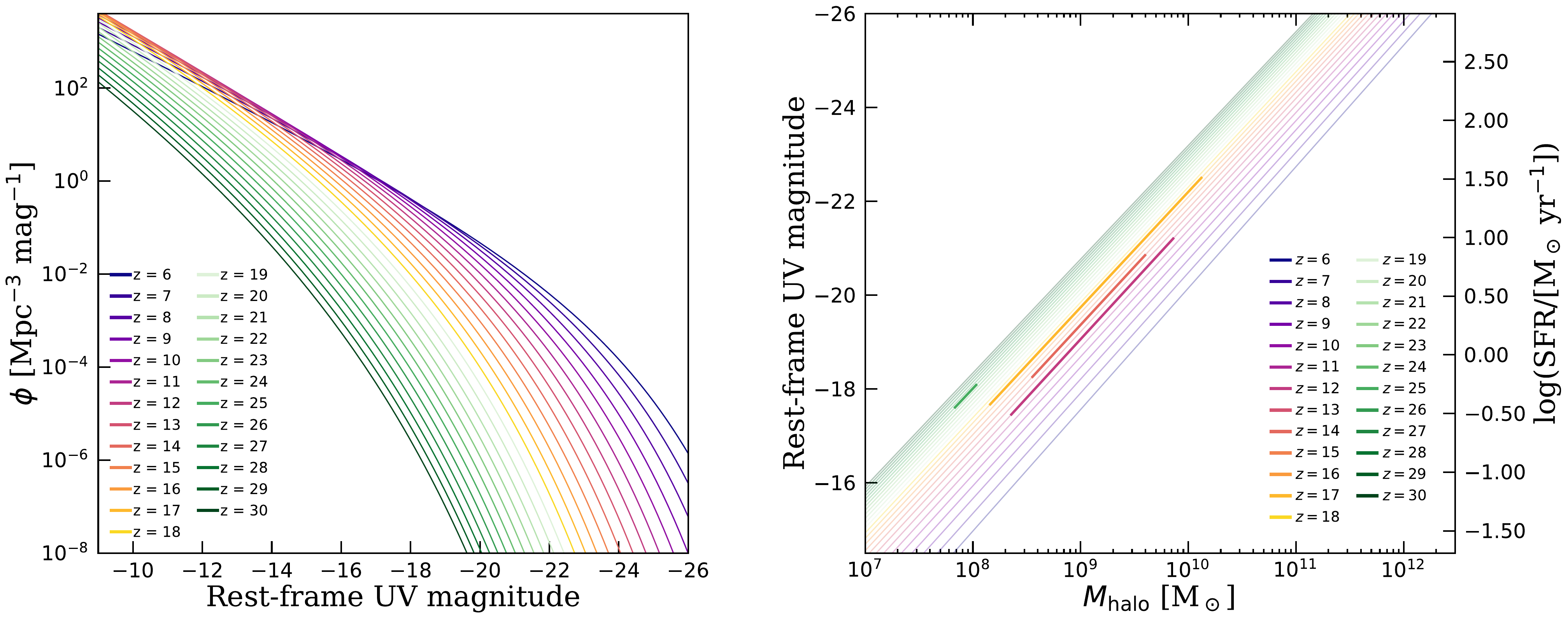}
    \caption{
        The \textit{left} panel shows predicted UV LFs between $z=6$ to 30 from our maximal $\epsilon_* = 1$ scenario. The \textit{right} panel shows the rest-frame FUV magnitudes and SFR as a function of $M_\text{halo}$ between $z=6$ to 30 (also in the $\epsilon_* = 1$ scenario), where the luminosity ranges (and their corresponding halo mass ranges) that are constrained by current observations are highlighted (see Section~\ref{sec:methods:obs} for details on the observational datasets).
    }
    \label{fig:UVLF_max_z30}
\end{figure*}

\section{Results}
\label{sec:results}

\begin{figure*}
    \includegraphics[width=2\columnwidth]{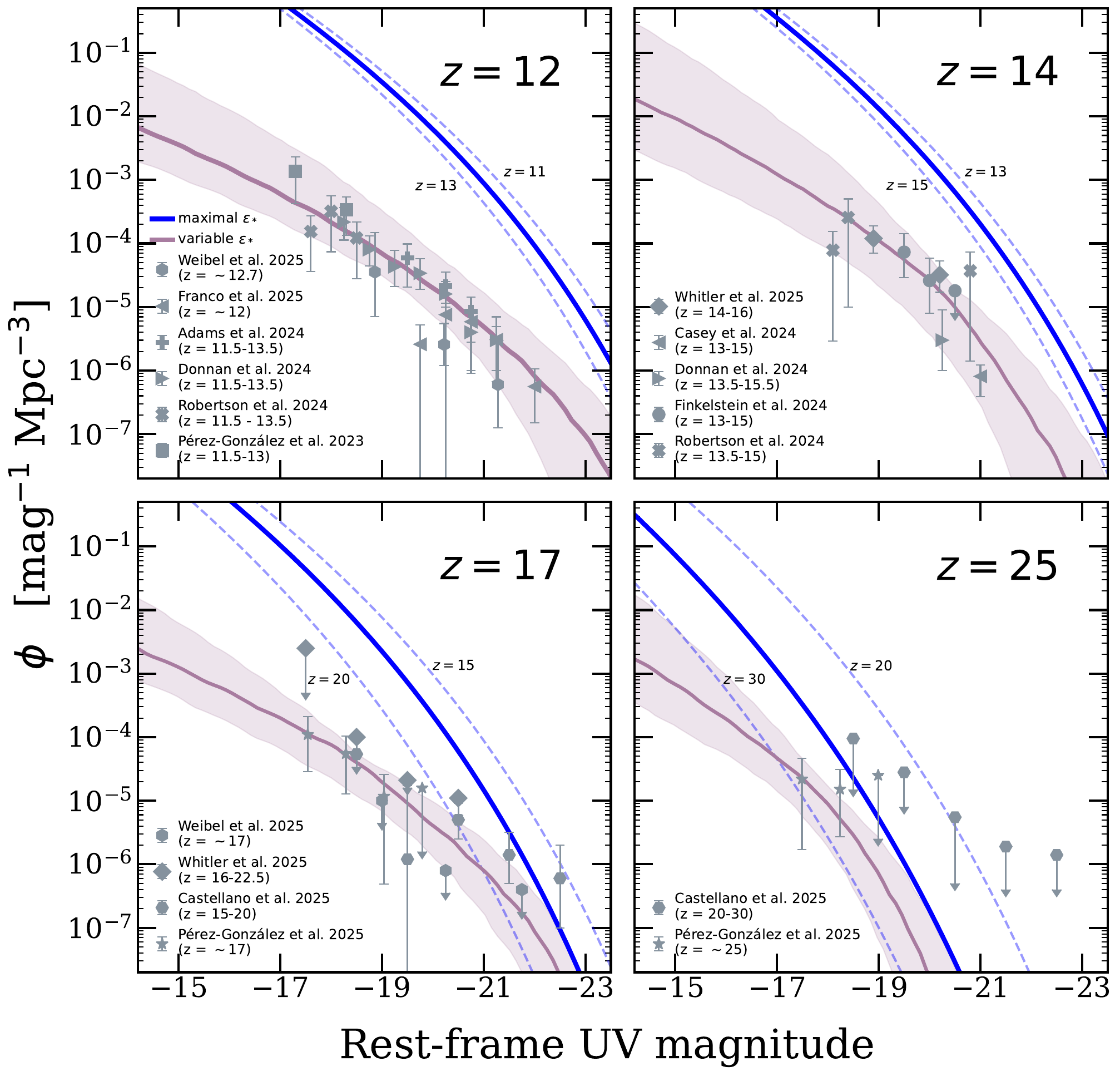}
    \caption{
        Simulated ultra-high-redshift rest-frame UV luminosity functions at $z=12$, 14, 17, and 25 assuming maximal $\epsilon_* = 1$ (blue solid line) and the posterior median and 16th and 84th percentile for a variable $\epsilon_*(M_\text{h},z)$ fitted to existing observations (purple line and shaded regions). These results are compared to ultra-high-redshift observational measurements reported by \citet{Donnan2024}, \citet{Adams2024}, \citet{Perez-Gonzalez2023}, \citet{Whitler2025}, \citet{Finkelstein2024}, \citet{Robertson2024}, \citet{Casey2024}, \citet{Castellano2025}, \citet{Perez-Gonzalez2025}, \citet{Franco2025}, and \citet{Weibel2025}. \emph{The fact that the blue lines are everywhere higher than the observations implies that there is no fundamental tension between $\Lambda$CDM and these observations}.
    }
    \label{fig:UVLF_max_obs}
\end{figure*}

Fig.~\ref{fig:UVLF_max_z30} shows the predicted UV luminosity function from $z\sim 6$--30 from our maximal model ($\epsilon_*=1$) in the left panel, and the implied relationship between halo mass and UV absolute magnitude under the assumption of the maximal model in the right panel. The darker lines provide a rough idea of the masses of the host halos that the observed galaxies must inhabit in order to match the observed number densities under the maximally efficient scenario. Fig.~\ref{fig:UVLF_max_obs} shows the UV luminosity function at redshifts $z\simeq 12$, 14, 17 and 25.  The maximal model prediction yields higher number densities at fixed UV magnitude at $z\sim 12$--14. At $z\sim 17$, the maximal model predicts number densities that lie above observed number densities at a given UV magnitude except for the brightest bin, for which it lies just within the 1$\sigma$ error bar. For $z\sim 25$, the maximal model is just above the brightest MIDIS+NGDEEP measurement ($M_{\rm UV} \sim -18)$ and about an order of magnitude above the measurement in the fainter bin. \emph{This implies that even if all of the photometric candidates are actually at these redshifts, there is no fundamental tension with $\Lambda$CDM.} Moreover, as we discuss further in Section~\ref{sec:discussion}, our `maximal' model is actually in some sense conservative, as bursty star formation and/or stellar populations with higher UV light-to-mass ratios (as might arise for lower metallicity or a top-heavy IMF) will boost the UVLF even further. In Fig.~\ref{fig:UVLF_max_obs}, we also show the UVLF from our model when we fit for the values of the four free parameters in the expression for $\epsilon_*(M_h, z)$ using MCMC as described in Section~\ref{sec:fitting}. We show the medians as well as the 1-$\sigma$ range of the posterior. 

\begin{figure}
    \includegraphics[width=\columnwidth]{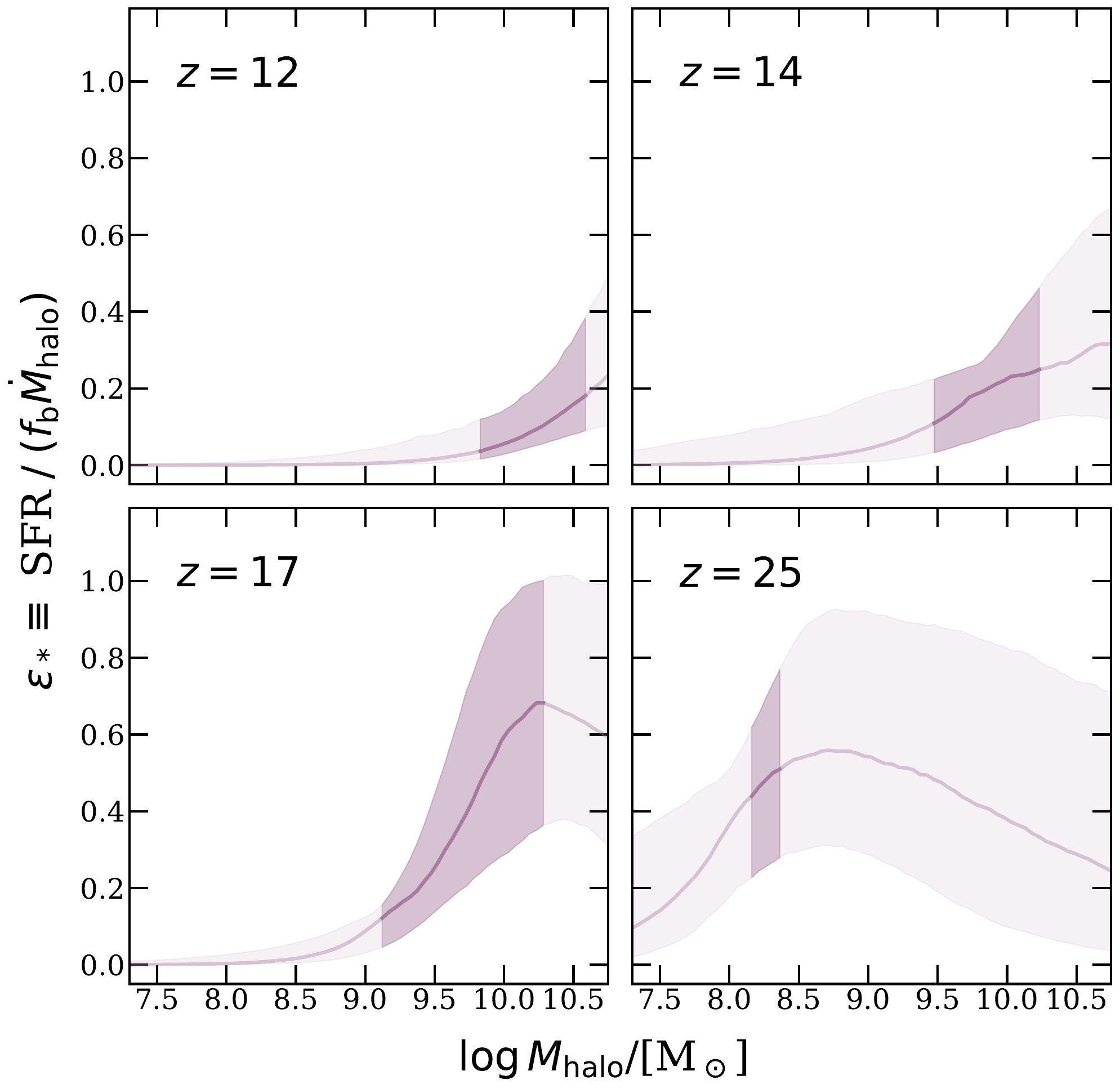}
    \caption{Median (solid lines) and 16th and 84th percentiles (shaded regions) for the values of the baryon conversion efficiency $\epsilon_{*}$ obtained from our MCMC fitting procedure. Darker lines and shaded regions show the approximate range of halo mass where there are current observational constraints, assuming the median relation for $\epsilon_{*}$. The maximum required values of $\epsilon_{*}$ where there are constraints range from $\sim 20$--60 \%, which is not in fundamental tension with $\Lambda$CDM.
    }
    \label{fig:eps_posterior}
\end{figure}

Figure~\ref{fig:eps_posterior} shows the posteriors of $\epsilon_*$ as a function of halo mass for $z\simeq 12$, 14, 17 and 25. Over the range of halo masses that is constrained by the observations for the median values of $\epsilon_*$ (shown as darker lines and shading), we find that the median values of the baryon conversion efficiencies peak at $\simeq 20$-- 65\%. Although significantly higher than is typical in the local universe, these efficiencies may not be implausible, as we argue in Section~\ref{sec:discussion}.  

\section{Discussion}
\label{sec:discussion}
\subsection{Caveats of our analysis}

\begin{figure}
    \includegraphics[width=\columnwidth]{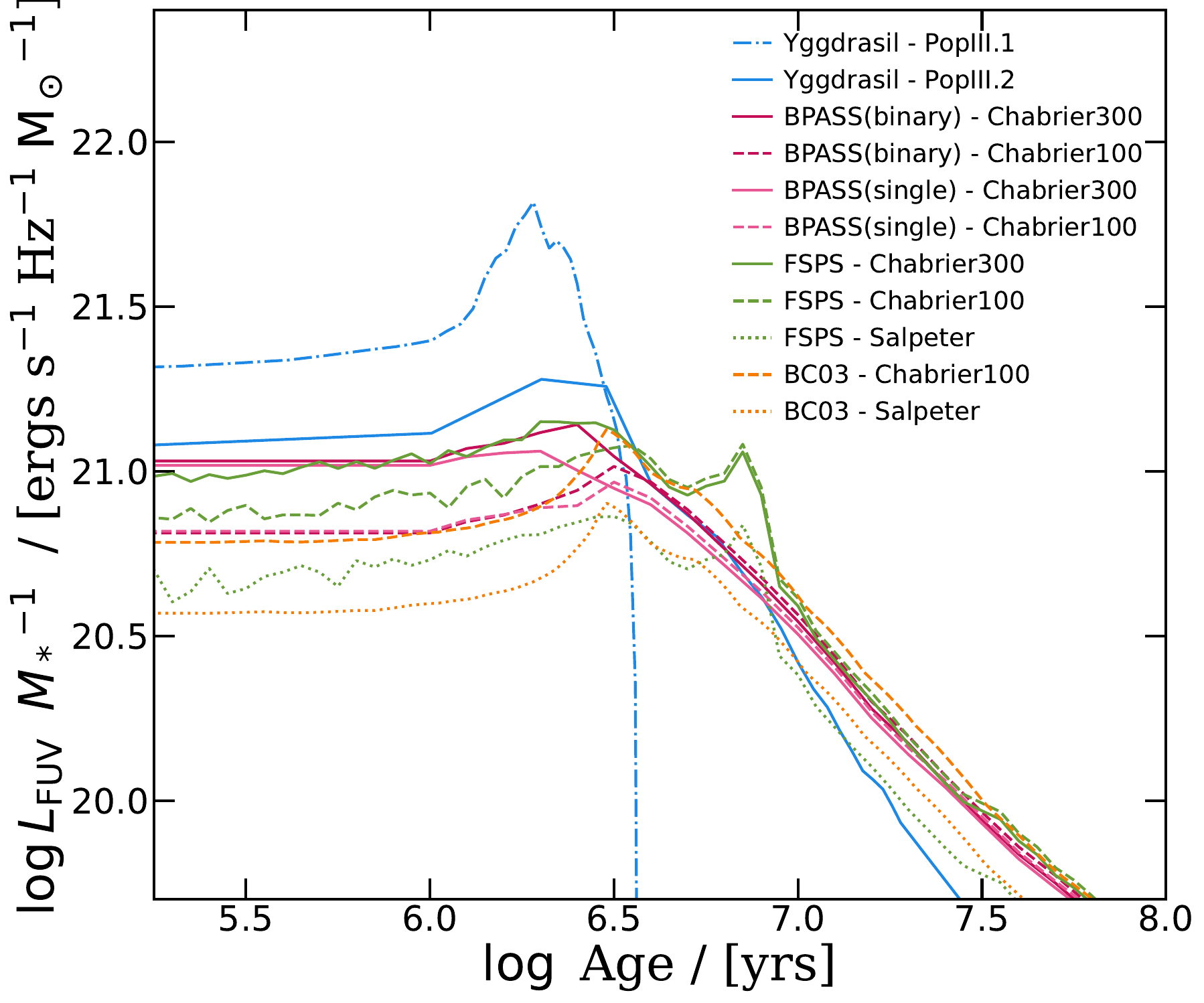}
    \caption{
        The far-UV luminosity $L_\text{FUV}$ per solar mass as a function of the age of a simple stellar population predicted by the \citetalias{Bruzual2003} (orange), \textsc{bpass} (red and pink for binary and single star model), \textsc{fsps} (green), and the Yggdrasil (blue) spectral synthesis models, assuming an instantaneous starburst at the lowest metallicity provided by these models, where $Z = 10^{-4}$, $10^{-5}$, and $2\times10^{-4}$, for \citetalias{Bruzual2003}, \textsc{bpass}, and \textsc{fsps}, respectively. We also indicate the IMF adopted by these models, with Chabrier300 and Chabrier100 indicating the use of a \citet{Chabrier2003a} IMF with upper mass cutoffs at 300\msun\ (solid lines) and 100\msun\ (dashed lines), respectively, and Salpeter indicating the use of a \citet{Salpeter1955} IMF (dotted lines). \textit{The choice of IMF has a significant impact on the predicted luminosity of young stellar populations, which is greater than the differences observed among various SPS models using the same IMF}.
    }
    \label{fig:LUV_Age}
\end{figure}

\begin{figure*}
    \includegraphics[width=2\columnwidth]{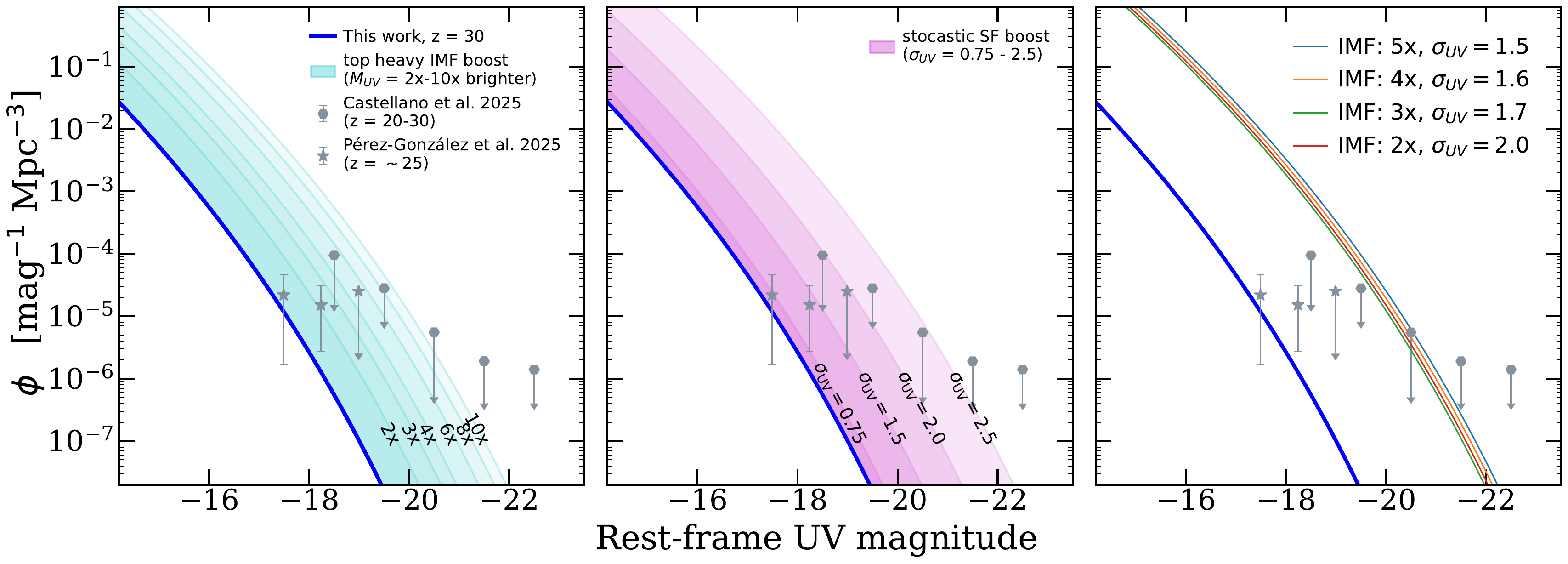}
    \caption{
        We explore the effect at $z=30$ of `boosted' UV luminosity due to a top heavy IMF \citep[similar to the approach adopted by][]{Yung2024} and UV variability due to stochastic star formation \citep{Shen2023a}. In the \textit{left panel}, the light blue shaded regions and lines show the effect of boosting the UV luminosity of all galaxies by a factor of 2--10. In the \textit{middle panel}, we show the effect of convolving the UV luminosity function with a Gaussian kernel of width, $\sigma_\text{UV}$, of 0.75 to 2.5. In the \textit{right panel}, we experiment with a handful of scenarios combining both boosting mechanisms, illustrating that a combination of a top-heavy IMF boost, similar to that predicted by Pop III stellar population synthesis models, and bursty star formation with a $\sigma_\text{UV}$ similar to that suggested by numerical simulations, can match the number density upper limits inferred by recently identified $z\sim 20$--30 galaxy candidates.
    }
    \label{fig:UVLF_z30_boost}
\end{figure*}

A significant uncertainty in our analysis --- probably the dominant one --- is the assumed  stellar light-to-mass ratio (SLMR). The SLMR for an integrated stellar population depends on the assumed IMF as well as the star formation and chemical enrichment history and the details of stellar evolution physics and stellar atmospheres that are baked into a given set of Simple Stellar Population (SSP) models \citep{Conroy2013, Iyer2025}. The MD14 conversion factor is based on an assumed chemical enrichment history (where a fixed metallicity is assigned to all stars in the galaxy, but this metallicity is a function of redshift) and a fixed star formation history, which at $z\gtrsim 6$ corresponds to a SFR that rises with time as a power law \citep{Madau2014}, convolved with the Flexible Stellar Population Synthesis \citep[\textsc{fsps}; ][]{Conroy2010} SSP models with a \citet{Salpeter1955} IMF. In our analysis we have performed a standard conversion to a Chabrier  IMF. Yung et al. (in prep) show that the MD14 conversion factor actually provides a remarkably good representation of the values of $\mathcal{K}_\text{UV}$ at $6 \lesssim z \lesssim 20$ that are obtained when the full star formation and chemical evolution histories predicted by a semi-analytic model of galaxy formation are convolved with SSP models. This lends support to our adoption of that assumption here. Furthermore, several known or expected effects would actually lead to brighter galaxies. Extremely low metallicity stellar populations, or a top-heavy IMF, would lead to light-to-mass ratios that can be higher than the MD14 baseline by as much as an order of magnitude, though this is for the relatively extreme assumption of a pure Pop III.1 (primordial) stellar population. More likely, the SLMR boost from an evolving IMF is of order a factor of a few \citep{Raiter2010,Chon2024}. Fig.~\ref{fig:LUV_Age} shows $L_{\rm UV}$ per 1 solar mass of stars for several different SSP models at low metallicity, including \citetalias{Bruzual2003} \citep{Bruzual2003}, \textsc{bpass} \citep{Eldridge2017, Stanway2018}, \textsc{fsps} \citep{Conroy2010}, and Pop III stars with Yggdrasil \citep{Zackrisson2011}. We note that $\mathcal{K}_{\rm UV}$ is degenerate with $\epsilon_0$, such that decreasing $\mathcal{K}_{\rm UV}$ by a given factor (making galaxies brighter for a given SFR) would decrease the implied values of $\epsilon_*$ by the same factor. 

Another assumption we have made is that there is no impact of dust attenuation on the emergent UV flux. This is supported by observational measurements of rest-UV slopes in galaxies at $11 \lesssim z \lesssim 14$, which are extremely blue and leave very little room for reddening by dust \citep{Cullen2023, Morales2024a}. However, standard supernova dust yields would actually suggest that even these early galaxies could already contain significant amounts of dust \citep{Ferrara2023}. Therefore it is possible that there is a population of galaxies that is extinguished by dust and are sufficiently dimmed that they drop out of the existing samples. Deeper observations should reveal whether there is a population of fainter galaxies with redder UV slopes. 

Our analysis so far furthermore assumes a deterministic relationship between halo mass and UV luminosity at any given redshift (through the halo accretion rate, which scales with mass and redshift). However, theoretical simulations predict that star formation in early galaxies is expected to be quite bursty \citep{Pallottini2023, Sun2023a}, and there is observational evidence that galaxies at $z \gtrsim 6$ experience bursty star formation \citep{Endsley2023a, Kokorev2025}. Bursty SF will cause a fraction of lower mass halos to be boosted in UV brightness. Because of the steep decline of the bright end of the observed UVLF, this effect leads to an Eddington bias that can significantly increase the number of luminous galaxies for a given underlying SFE, as has been shown by several previous studies based on empirical models similar to ours, but with a stochastic SF component added \citep{Shen2023a, Gelli2024, Shuntov2025}. Thus incorporating a bursty SF component would lead to lower inferred SFE in our analysis. Because of the large uncertainties in the observed UVLF at $z\gtrsim 15$ and the very degenerate parameter space, we do not attempt to incorporate a bursty component in the full posteriors presented here. We note that observational constraints on galaxy clustering could in principle help break the degeneracy between burstiness and globally increased star formation efficiency \citep{Munoz2023,Sun_clustering2025}.

In Fig.~\ref{fig:UVLF_z30_boost}, we illustrate 
how variations in SLMR and burstiness could impact our predictions at the highest redshifts we have considered.  Similar to the approach presented in fig.~4 of \citet{Yung2024}, we show the impact on the predicted $z\sim30$ UV luminosity function if the luminosity-to-mass ratio of stellar populations is boosted by a factor of 2 to 10, which could occur due to a top-heavy IMF \citep{Raiter2010}. Similarly, we show the impact on the same galaxy populations from adding a scatter in UV luminosity at fixed halo mass, $\sigma_\text{UV}$, due to stochastic starbursts, which may occur on time-scales shorter than one halo dynamical time and temporarily boost the UV luminosity. It is evident that these boosts can significantly increase the number of bright galaxies predicted at such extreme redshifts, even potentially accommodating the most optimistic view of the \citet{Perez-Gonzalez2025} and \citet{Castellano2025} candidates.

While a factor of 10 boost from a top-heavy IMF alone or a scatter $\sigma_\text{UV} \sim 2.5$ may seem extreme, the combined effect of these mechanisms can yield significant enhancements in galaxy number densities. As shown in Fig.~\ref{fig:LUV_Age}, a PopIII.1 (first generation Pop III with characteristic masses of $\sim 100$ \msun) stellar population can be a factor of $\sim 2.5$ to 8 more luminous than a metal-poor stellar population of comparable age. As shown in the right panel of Fig.~\ref{fig:UVLF_z30_boost}, combining this with a $\sigma_\text{UV}$ similar to that predicted by high-resolution numerical hydrodynamic zoom-in simulations \citep[e.g.][]{Sun2023a} can boost the UV luminosity functions to match the $M_\text{UV}\lesssim-20$ upper limits at $z\sim 20$ to 30 presented by \citet{Castellano2025} and exceed the observed number density at fainter magnitudes.

Another source of uncertainty is the possible contribution of UV light from AGN to the total observed galaxy UV luminosity. The expected contribution from SMBH arising from `standard' seeding models (i.e. Pop III or direct collapse seeds) is not expected to be large \citep{Trinca2024}, simply because BH massive enough to contribute a significant amount of UV light are expected to be rare at these epochs (and moreover, rapidly accreting BH are likely to be heavily obscured in the UV). \citet{Matteri2025, Matteri2025a} propose that a population of primordial black holes with masses $10^4$--$10^5$ \msun\ hosted by low mass halos ($M_{\rm h} \sim 10^7$ \msun), accreting at about a third of their Eddington luminosity, could provide an alternative explanation for the unexpectedly large number density of UV luminous galaxies at $z\gtrsim 10$. They suggest that this scenario can be tested observationally with spectroscopic followup with JWST.

To conclude this sub-section, the observational measurements themselves have very large uncertainties. As discussed in detail in \citet{Castellano2025} and \citet{Perez-Gonzalez2025}, it is very possible --- even likely --- that at least some (and perhaps all) of the photometrically selected $15 \lesssim z \lesssim 30$ candidates are actually interlopers at much lower redshifts. Currently, the observations that put the most strain on models are the three F200W dropouts presented in \citet{Castellano2025}, with very bright implied rest-UV luminosities ($M_{\rm UV} < -21$). These objects all have strongly multi-modal redshift probability distribution functions (PDF; see Fig.~3 of \citealt{Castellano2025}), with a significant peak at lower redshift ($z \sim 5$), and may be similar to known $z\sim 16$ `impostors' which spectroscopic follow-up revealed to actually be at $z \sim 5$ \citep{ArrabalHaro2023}. These objects are bright enough that they should be relatively easy to follow up spectroscopically.  The second highest tension comes from the three F277W dropouts reported as $20 \lesssim z \lesssim 30$ galaxy candidates by \citet{Perez-Gonzalez2025}. These objects do not have strongly multi-modal redshift PDFs, but they are only detected in two photometric filters. These objects are extremely faint ($m_{\rm AB} \gtrsim 30$--31) and will be challenging to follow up with spectroscopy with any currently available facilities. However, studies of lensed cluster fields may yield candidates with similar intrinsic luminosities that can be followed up spectroscopically. As pointed out by \citet{Gandolfi2025}, lower-mass, lower redshift dust-obscured galaxies can also have similar colors to $z>15$ galaxies.

\subsection{Implications of our results}

Even if all of the $z\gtrsim 15$ galaxy candidates are really at these extreme high redshifts, our analysis has shown that they can be explained without discarding the $\Lambda$CDM framework or invoking `exotic' mechanisms such as accreting primordial black holes. However, explaining them within this `vanilla' framework does require invoking baryon conversion efficiencies that are significantly higher than in the local universe. Are the values of baryon conversion efficiencies implied by our analysis (which peak at values of $0.2$--0.65) physically plausible? And if so, what physical processes could lead to efficiencies that are so much higher than those in the local universe?

The first stars are thought to form via primordial H$_2$ cooling in halos with $M_{\rm h} \gtrsim 10^5$ to a few $\times 10^6$\msun\ at a redshift of $z\sim 30$, and the minimum halo mass for which gas can cool via atomic cooling ($T_{\rm vir} > 10^4$ K) ranges from a few times $10^7$ \msun\ at $z\sim 15$ to $\sim 5 \times 10^6$ \msun\ at $z\sim 30$ \citep{Klessen2023}. The cooling times and dynamical times at these epochs are extremely short, of the order of Myr. Thus cooling and accretion are not expected to be the rate limiting factors for star formation in this halo mass and redshift regime. The physical processes that regulate the conversion of baryons into stars once gas has cooled into the ISM and formed Giant Molecular Clouds (GMC) are thought to be winds, jets, photoionizing radiation and other radiation from young massive stars on cloud scales, and supernova driven winds on large (galactic) scales \citep{Chevance2023,Hopkins2012}. In the local Universe, the cloud scale star formation efficiency per free-fall time is a few percent \citep{krumholz2007}, and the global integrated galaxy scale baryon conversion efficiency $m_*/(f_b M_{\rm h})$ is a few percent (or even less) in low-mass halos, and peaks at a value of around 20\% in halos with masses of $\sim 10^{12}$\msun\, \citep{Moster2010,Behroozi2013d,Behroozi2019}. 

Ultra-high redshift ($z\gtrsim 10$) galaxies are extremely compact and dense, with star formation surface densities of tens to hundreds of \msun/yr/kpc$^2$, exceeding the corresponding values in local galaxies by up to four orders of magnitude \citep{Morishita2024}. If the individual star forming clouds within these galaxies are similarly boosted in density, their free-fall times will be extremely short, of order just a few Myr. Under these conditions, star formation is expected to be regulated largely by the processes associated with the massive stars on cloud scales, as the time for the first supernovae to explode is longer than the lifetime of these clouds \citep{Dekel2023}. Numerical simulations of star formation on GMC scales have shown that clouds with higher surface density are able to convert much larger fractions of their gas into stars over the cloud lifetime \citep{Lancaster2021b,Polak2024,Menon2024a}, with overall cloud-scale efficiencies as high as 90\% for clouds with surface densities of $10^4$--$10^5$ \msun. The global SFE will clearly depend on the fraction of the ISM that is in these dense clouds, which is unknown. However, we note that values as high as 50\% are supported by observations of lensed systems at $z\sim 6$--10  \citep{Adamo2024,Fujimotograpes2024}.

\subsection{Comparison with Results from the Literature}

\citet{Somerville2025} incorporated density-modulated star formation efficiencies, motivated by the GMC-scale simulations mentioned above, into a semi-analytic galaxy formation model, under the assumption that the cloud surface densities are comparable to the overall gas surface density and the galaxy size scales with the halo virial radius. When they assume that about half of the total ISM mass is in dense star forming clouds, their model reproduces the UV luminosity function of the fainter $z\sim 17$ candidates reported by \citet{Perez-Gonzalez2025}, while to reproduce the bright $z\sim 17$ candidates reported by \citet{Castellano2025}, they required 100\% of the ISM to be in dense star forming clouds (predictions from this model are not currently available for $z \gtrsim 17$). \citet{Boylan-Kolchin2025} has suggested a similar picture, in which the high density and gravitational acceleration of early dark matter halos makes stellar feedback inefficient, and star formation more efficient. Our conclusions are also consistent with the work based on semi-analytic models by \citet{Mauerhofer2025}, who concluded that either a time evolving IMF or a time evolving SFE is needed to account for the observed UV LFs at $z\gtrsim 11$.

\citet{Toyouchi2025} employed an empirical method similar to ours and derived $\epsilon_*$ constraints between $5 \lesssim z \lesssim 20$  from the cosmic star formation density inferred by \textit{JWST} observations, which are broadly consistent with our findings. Similarly, a Bayesian analysis of \textit{Hubble} and \textit{JWST} UVLFs presented by \citet{Dhandha2025} between $6\lesssim z \lesssim 14.5$ yield halo-mass and redshift-dependent SFE constraints that are consistent with our results.

To date, there are no numerical galaxy scale simulations that fully resolve the interplay between massive stars and star forming gas on GMC scales, so star formation prescriptions must be treated via sub-grid models in these simulations. In many/most cases (particularly in large volume simulations), the star formation efficiencies on both cloud and galaxy scales are either hard-wired or tuned to be comparable to the (low) values in the local universe. A few recent high-resolution simulations that attempt to implement star formation prescriptions in a more `first-principles' manner find star formation efficiencies of up to  $\sim 20$ \% at $z\sim 9$--13, significantly higher than the canonical local universe values \citep{Ceverino2024,Andalman2024}. 

Our results are in qualitative agreement with the hydrodynamic simulations of \citet{Semenov2024}, which found very high global star formation efficiencies despite low ($\sim 1\%$) average local efficiencies per free-fall time, driven primarily by the very high densities of star-forming gas. For redshifts up to $z \sim 14$, FIREbox$^{HR}$ \citep{Feldmann2025}, a high-resolution cosmological hydrodynamical simulation, and \textsc{thesan-zoom} \citep{Shen2025}, a zoom-in radiation-hydrodynamic simulation both reported $\epsilon_* \lesssim 1\%$. \citet{Feldmann2025} found that in halos with $M_\text{h} \lesssim 10^{10}$ \msun, an efficiency of $\epsilon_* \lesssim 1\%$ in FIREbox$^{HR}$ is sufficient to reproduce the faint-end slope of the observed UVLFs extrapolated to magnitudes below the current detection limit. In contrast, the \textsc{thesan-zoom} simulations suggest that the tension with the high number of luminous galaxies detected by \textit{JWST} can be alleviated by both a higher SFE than reported in pre-\textit{JWST} studies and by bursty star formation.
Our inferred $\epsilon_*$ at $z \sim 12$ is in agreement with these results over the halo mass and redshift ranges probed by these simulations, given their limited volume and mass resolution.

\section{Conclusions}
\label{sec:conclusions}

In this work, we address whether the recently announced extreme ultra-high-z galaxy candidates at $15 \lesssim z \lesssim 30$ \citep{Perez-Gonzalez2025,Castellano2025} would pose fundamental tension with the standard $\Lambda$CDM paradigm or require exotic explanations, if they are actually at these very high redshifts. 
We have made use of the recent \gureft\ suite of dissipationless cosmological $N$-body simulations, which provide predictions of dark matter halo mass functions and growth rates from $6 \lesssim z \lesssim 30$ for state-of-the-art cosmological parameters at unprecedented precision over a wide range of halo masses ($5 \lesssim \log(M_h/$\msun$) \lesssim 12.5$). We combined these results with a simple empirical model that posits that the SFR is proportional to the inflow rate of baryons into the halo, times an adjustable efficiency factor $\epsilon_*(M_h, z)$. Our main results are as follows:

\begin{itemize}
\item We provide convenient fitting functions for dark matter halo mass functions and halo growth rates from $6 \lesssim z \lesssim 30$. We find that these robust results adopting state-of-the-art constraints on cosmological parameters can differ significantly from commonly used halo mass functions in the literature using out of date cosmological parameters \citep{Reed2007} and analytic approximations \citep{Sheth1999}.

\item We find that, with conservative assumptions about the UV light-to-mass ratio $L_{\rm UV}/m_*$ for stellar populations, even if \emph{all} of the existing $15 \lesssim z \lesssim 30$ candidates are actually at these extreme high redshifts, \emph{there is no fundamental tension with $\Lambda$CDM and no need for exotic sources}.

\item Using our simple modeling framework, and again adopting conservative assumptions about $L_{\rm UV}/m_*$, we find required peak baryon conversion efficiencies of 0.2--0.65. These are consistent with values seen in numerical simulations of the dense gas clouds that are probably typical at these epochs. 

\item Our analysis is conservative, as we have not accounted for bursty star formation, which is likely to be common in early galaxies and will lead to even larger numbers of UV bright galaxies. In addition, it is entirely plausible that early stellar populations could have higher $L_{\rm UV}/m_*$ than what we have assumed, due to lower metallicities and/or a top-heavy IMF. These effects could plausibly reduce the required star formation efficiencies by a factor of a few. 

\end{itemize}

More deep observations, both imaging and spectroscopy, with JWST are needed to determine whether such extreme high redshift galaxies are common in other parts of the sky, and to identify and better understand lower redshift interlopers. In addition, more work with high resolution numerical simulations including detailed physics-based treatments of star formation and stellar feedback is needed to understand the plausibility and physical origin of the apparently higher star formation efficiencies in early galaxies. 

\section*{Data Availability}
The data used in this analysis will be made available upon request. 

\section*{Acknowledgments}
We warmly thank the CCA galaxy formation group for helpful suggestions that improved the presentation of this work.
We also thank the anonymous referee for the constructive comments that improved this work.
The analysis in this work was carried out with \textsc{astropy} \citep{Robitaille2013, Price-Whelan2018}, \text{pandas} \citep{Reback2022}, \text{numpy} \citep{VanderWalt2011}, and \textsc{scipy} \citep{Virtanen2020}.
We list here the roles and contributions of the authors according
to the Contributor Roles Taxonomy (CRediT).\footnote{\url{https://credit.niso.org}} L. Y. Aaron Yung: conceptualization, data curation, formal analysis, visualization, methodology, and editing. Rachel S. Somerville: conceptualization, data curation, methodology, and writing original draft. Kartheik Iyer: formal analysis, methodology, and visualization.
The \gureft\ simulation suite was run on the computing cluster managed by the Scientific Computing Core (SCC) of the Flatiron Institute.
AY is supported by a Giacconi Fellowship from the Space Telescope Science Institute, which is operated by the Association of Universities for Research in Astronomy, Incorporated, under NASA contract HST NAS5-26555 and JWST NAS5-03127. The Flatiron Institute is supported by the Simons Foundation. 
Support for KI was provided by NASA through the NASA Hubble Fellowship grant HST-HF2-51508 awarded by the Space Telescope Science Institute.
This work is based on observations made with the NASA/ESA/CSA James
Webb Space Telescope, obtained at the Space Telescope Science Institute, which
is operated by the Association of Universities for Research in Astronomy, Incorporated,
under NASA contract NAS5-03127.




\bibliographystyle{mnras}
\bibliography{ultra-z-empirical} 




\appendix

\section{Comparison with other HMF fitting functions}
\label{sec:appendixA}
In this Appendix, we compare our ultra-high-redshift HMF fitting functions, fitted to halos extracted from our \gureft\ suite \citepalias{Yung2024a} and the Very Small MultiDark Planck simulation \citep{Klypin2016}, with other commonly used HMF fitting functions. A similar comparison is presented in \citetalias{Yung2024a} out to a maximum redshift of $z=15$. This comparison includes the analytic model from \citet{Sheth2002}, high-resolution $N$-body simulations of \citet{Reed2007}, which were carried out with earlier WMAP3 cosmological parameters \citep{Spergel2007}, and the \citet{Tinker2008} fitting functions implemented by the \textsc{hmf}\footnote{\url{https://hmf.readthedocs.io}, version 3.5.1} \citep{Murray2013} and the \textsc{colossus} \citep{Diemer2018} packages. We note that the HMFs produced by these packages are sensitive to the details of the implementation and are thus included in this comparison.

We also note that we are adapting the same functional form as the one implemented by \citet{Rodriguez-Puebla2016}. However, one of their fitting parameters changes sign at $z\sim12.63$ and their results cannot be sensibly extrapolated beyond that.

\begin{figure}
    \includegraphics[width=\columnwidth]{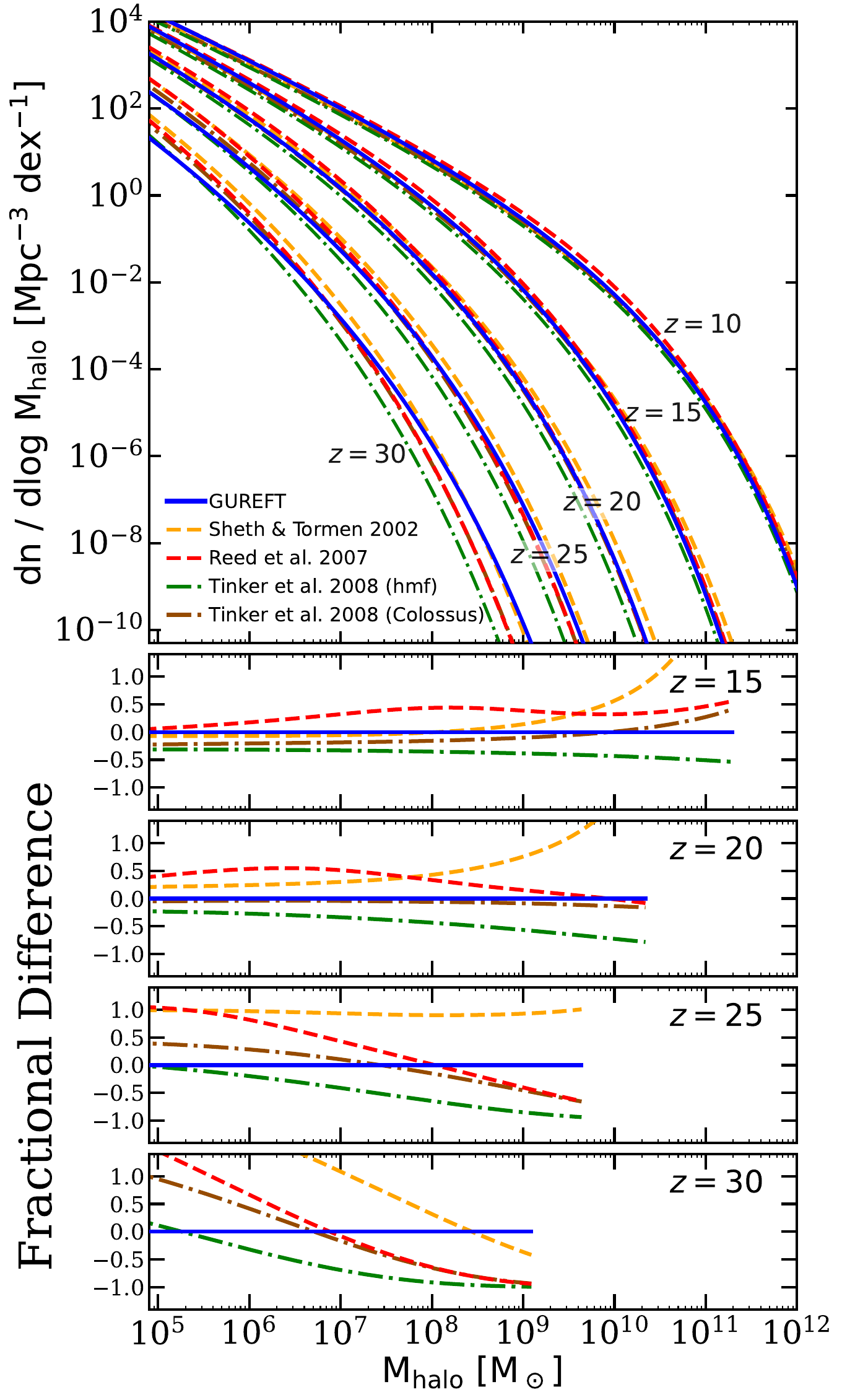}
    \caption{
        The HMF from our \gureft+Bolshoi-Planck based fitting functions (blue) compared with fitting functions and an analytic model from the literature (references shown on figure panel) at $z = 10$, 15, 20, 25, and 30. The bottom panels show the fractional difference of the number density of halos from these HMFs relative to our fitting functions ($\phi_X - \phi_\text{GUREFT+BP})/\phi_\text{GUREFT+BP}$) at $z > 10$. These commonly used HMF can differ from our state-of-the-art, high precision results by up to an order of magnitude, and the discrepancies vary with halo mass and redshift. 
    }
    \label{fig:hmf_compare_ultraz}
\end{figure}

\section{Posteriors for star formation efficiency parameters}
\label{sec:appendixB}

In this Appendix, we present the posteriors for the $\epsilon_*$ parameters for equation \ref{eqn:epsilon_star} obtained from our MCMC fitting procedure presented in Section \ref{sec:fitting}.

\begin{figure*}
    \includegraphics[width=0.9\columnwidth]{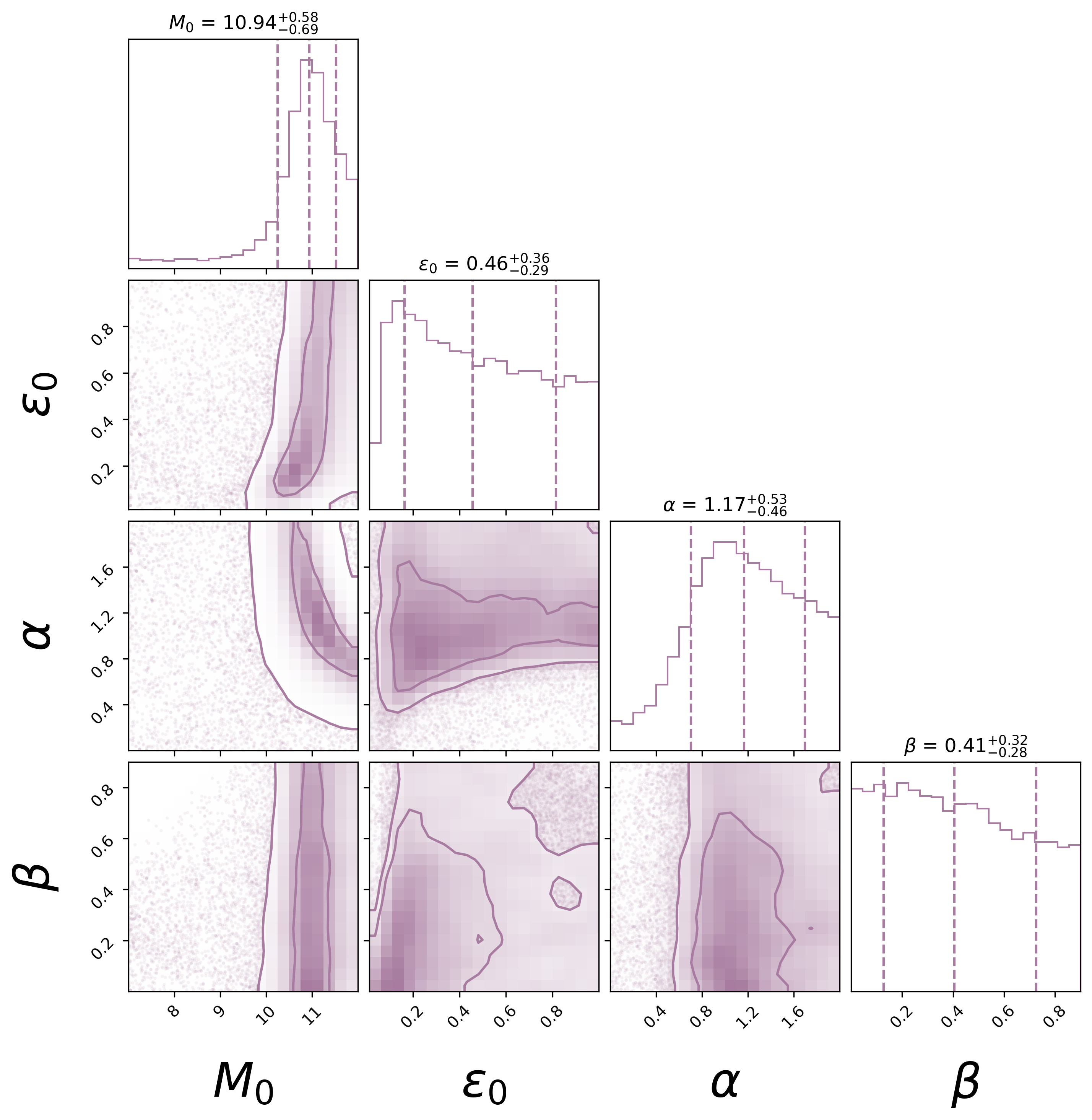}
    \includegraphics[width=0.9\columnwidth]{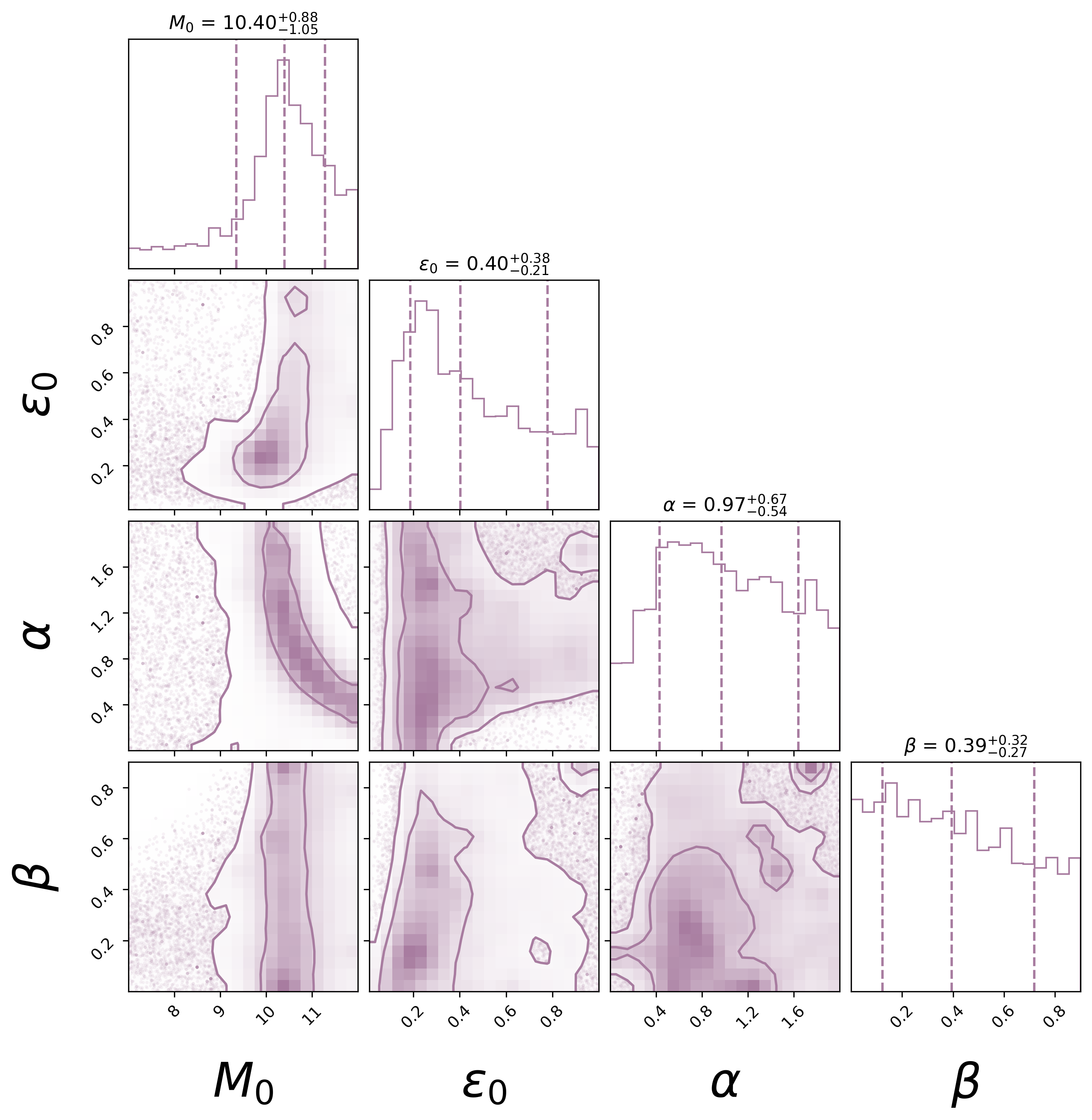}
    \includegraphics[width=0.9\columnwidth]{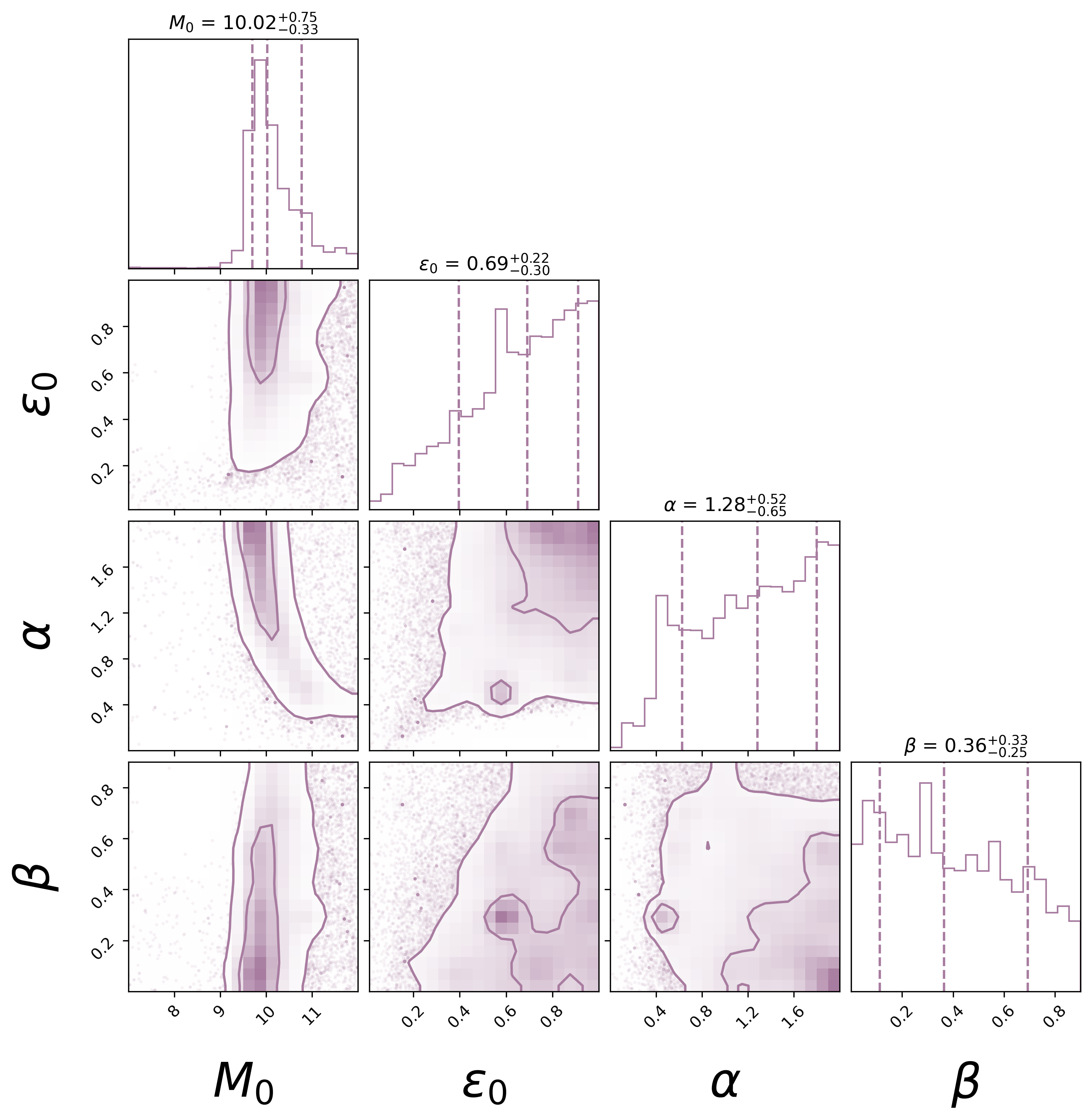}
    \includegraphics[width=0.9\columnwidth]{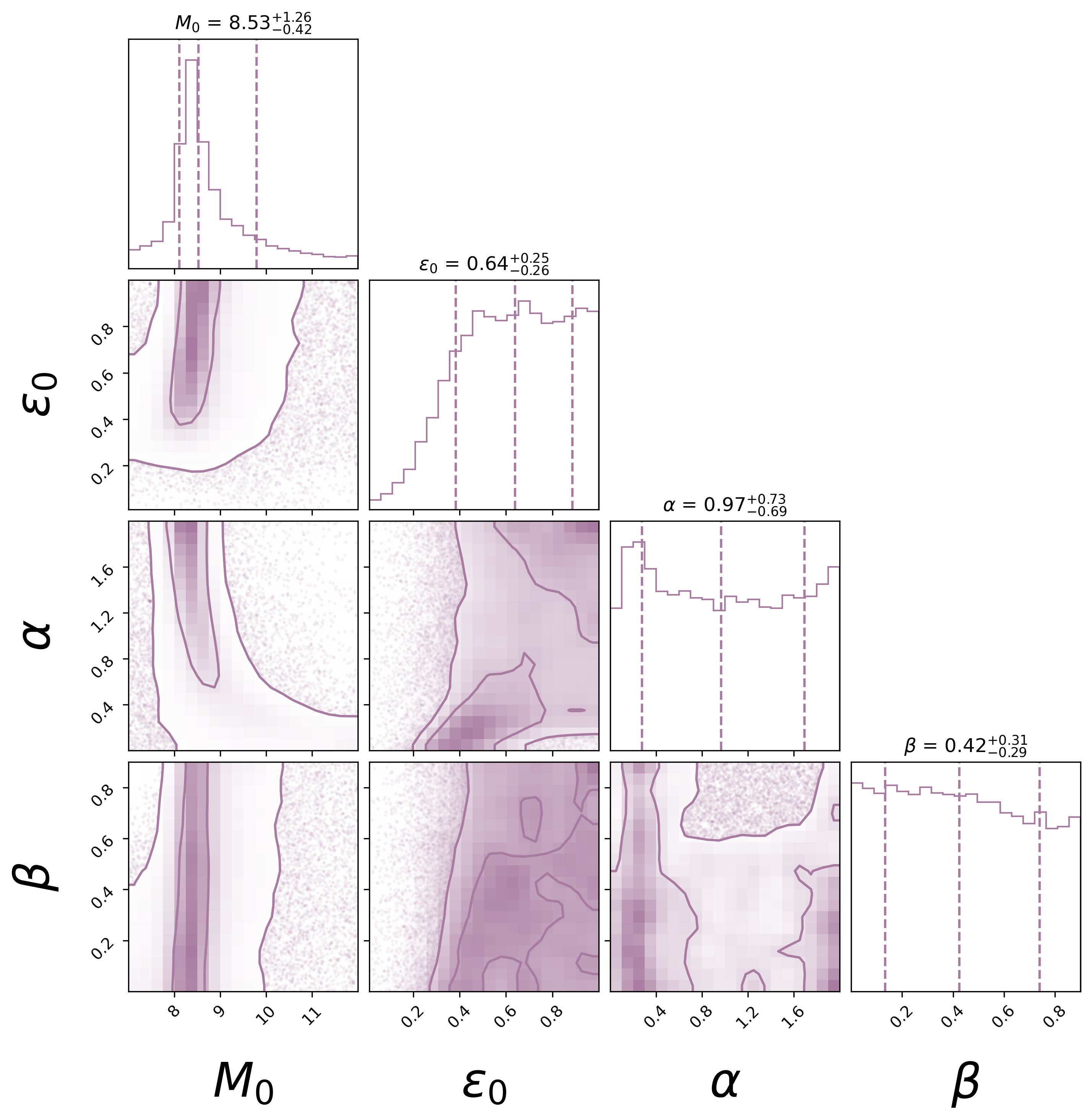}
    \caption{
        Parameter space posteriors for the four parameters of our empirical model ($\epsilon_0$, $M_0$, $\alpha$ and $\beta$) at $z = 12$ \textit{(top-left)}, 14 \textit{(top-right)}, 17 \textit{(bottom-left)}, and 25 \textit{(bottom-right)}.
    }
    \label{fig:efficiency_param_posteriors}
\end{figure*}


\bsp	
\label{lastpage}
\end{document}